# *k*-fingerprinting: a Robust Scalable Website Fingerprinting Technique

Jamie Hayes
*University College London*
`j.hayes@cs.ucl.ac.uk`

George Danezis
*University College London*
`g.danezis@ucl.ac.uk`

## Abstract

Website fingerprinting enables an attacker to infer which web page a client is browsing through encrypted or anonymized network connections. We present a new website fingerprinting technique based on random decision forests and evaluate performance over standard web pages as well as Tor hidden services, on a larger scale than previous works. Our technique, *k*-fingerprinting, performs better than current state-of-the-art attacks even against website fingerprinting defenses, and we show that it is possible to launch a website fingerprinting attack in the face of a large amount of noisy data. We can correctly determine which of 30 monitored hidden services a client is visiting with 85% true positive rate (TPR), a false positive rate (FPR) as low as 0.02%, from a world size of 100,000 unmonitored web pages. We further show that error rates vary widely between web resources, and thus some patterns of use will be predictably more vulnerable to attack than others.

## 1 Introduction

Traditional encryption obscures only the content of communications and does not hide metadata such as the size and direction of traffic over time. Anonymous communication systems obscure both content and metadata, preventing a passive attacker from observing the source or destination of communication.

Anonymous communications tools, such as Tor [11], route traffic through relays to hide its ultimate destination. Tor is designed to be a low-latency system to support interactive activities such as instant messaging and web browsing, and does not significantly alter the shape of network traffic. This allows an attacker to exploit information leaked via the order, timing and volume of resources requested from a website. As a result, many works have shown that website fingerprinting attacks are possible even when a client is browsing with encryption or using an anonymity tool such as Tor [7, 16, 17, 21, 23, 27, 32, 36, 38, 39].

Website fingerprinting is commonly formulated as a classification problem. An attacker wishes to know whether a client browses one of *n* web pages. The attacker first collects many examples of traffic traces from each of the *n* web pages by performing web-requests through the protection mechanism under attack; features are extracted and a machine learning algorithm is trained to classify the website using those features. When a client browses a web page, the attacker passively collects the traffic, passes it in to their classifier and checks if the client visited one of the *n* web pages. In the literature this is referred to as the closed-world setting – a client is restricted to browse a limited number of web pages, monitored by the attacker. However, the closed-world model has been criticized for being unrealistic [17, 29] since a client is unlikely to only browse a limited set of web pages. The open-world setting attempts to model a more realistic setting where the attacker monitors a small number of web pages, but allows a client to additionally browse to a large world size of unmonitored web pages.

Our attack is based on random decision forests [6], an ensemble method using multiple decision trees. We extend the random forest technique to allow us to extract fingerprints to perform identification in an open-world.

The key contributions of this work are as follows:

- A new attack, *k*-fingerprinting, based on extracting a fingerprint for a web page via a novel variant of random forests. We show *k*-fingerprinting is more accurate and faster than other state-of-the-art website fingerprinting attacks [7, 28, 39] even under proposed website fingerprinting defenses.
- An analysis of the features used in this and prior work to determine which yield the most information about an encrypted or anonymized web page. We show that simple features such as counting the number of packets in a sequence leaks more information about the identity of a web page than complex features such as packet ordering or packet inter-arrival time features.
- An open-world setting that is an order of magnitude

larger than the previous open-world website fingerprinting work of 5,000 unmonitored web pages [39][1], and nearly twice as large in terms of unique numbers websites than [28], reflecting a more realistic website fingerprinting attack over multiple browsing sessions. In total we tested *k*-fingerprinting on 101,130 unique websites[2].
- We show that a highly accurate attack can be launched by training a small fraction of the total data, greatly reducing the start-up cost an attacker would need to perform the attack.
- We observe that the error rate is uneven and so it may be advantageous to throw away some training information that could confuse a classifier. An attacker can learn the error rate of their attack from the training set, and use this information to select which web pages they wish to monitor in order to minimize their error rates.
- We confirm that browsing over Tor does not provide any additional protection against fingerprinting attacks over browsing using a standard web browser. Furthermore we show that *k*-fingerprinting is highly accurate on Tor hidden services, and that they can be distinguished from standard web pages.

## 2 Related Work

**Website Fingerprinting.** Website fingerprinting has been studied extensively. Early work by Wagner and Schneier [34], Cheng and Avnur [10] exposed the possibility that encrypted HTTP GET requests may leak information about the URL, conducting preliminary experiments on a small number of websites. They asked clients in a lab setting to browse a website for 5-10 minutes, pausing two seconds between page loading. With caching disabled they were able to correctly identify 88 pages out of 92 using simple packet features. Early website fingerprinting defenses were usually designed to safeguard against highly specific attacks. In 2009, Wright et al. [40] designed 'traffic morphing' that allowed a client to shape their traffic to look as if it was generated from a different website. They were able to show that this defense does well at defeating early website fingerprinting attacks that heavily relied on exploiting unique packet length features [21, 32].

In a similar fashion, Tor pads all packets to a fixed-size cells of 512 bytes. Tor also implemented randomized ordering of HTTP pipelines [30] in response to the attack by Panchenko et al. [27] who used packet ordering features to train an SVM classifier. This attack on Tor achieved an accuracy of 55%, compared to a previous attack that did not use such fine grained features achieving 3% accuracy on the same data set using a Naive-Bayes classifier [16]. Other defenses such as the decoy defense [27] loads a camouflage website in parallel to a legitimate website, adding a layer of background noise. They were able to show using this defense attack accuracy of the SVM again dropped down to 3%.

Luo et al. [24] designed the HTTPOS fingerprinting defense at the application layer. HTTPOS acts as a proxy accepting HTTP requests and obfuscating them before allowing them to be sent. It modifies network features on the TCP and HTTP layer such as packet size, packet time and payload size, along with using HTTP pipelining to obfuscate the number of outgoing packets. They showed that HTTPOS was successful in defending against a number of classifiers [5, 9, 21, 32].

More recently Dyer et al. [12] created a defense, BuFLO, that combines many previous countermeasures, such as fixed packet sizes and constant rate traffic. Dyer et al. showed this defense improved upon other defenses at the expense of a high bandwidth overhead. Cai et al. [8] made modifications to the BuFLO defense based on rate adaptation again at the expense of a high bandwidth overhead. Following this Nithyanand et al. [25] proposed Glove, that groups website traffic into clusters that cannot be distinguished from any other website in the set. This provides information theoretic privacy guarantees and reduces the bandwidth overhead by intelligently grouping web traffic in to similar sets.

Cai et al. [7] modified the kernel in Panchenko et al.'s SVM to improve an attack on Tor, and was further improved in an open-world setting by Wang and Goldberg in 2013 [38], achieving a true positive rate (TPR) of over 0.95 and a false positive rate (FPR) of 0.002 when monitoring one web page. Wang et al. [39] conducted attacks on Tor using large open-world sets. Using a *k*-nearest neighbor classifier they achieved a TPR of 0.85 and FPR of 0.006 when monitoring 100 web pages out of 5100 web pages. More recently Wang and Goldberg [37] suggested a defense using a browser in half-duplex mode – meaning a client cannot send multiple requests to servers in parallel. In addition to this simple modification they add random padding and show they can even foil an attacker with perfect classification accuracy with a comparatively (to other defenses) small bandwidth overhead. Wang and Goldberg [36] then investigated the practical deployment of website fingerprinting attacks on Tor. By maintaining an up-to-date training set and splitting a full packet sequence in to components comprising of different web page load traces they show that practical website fingerprinting attacks are possible. By considering a time gap of 1.5 seconds between web page loads, their splitting algorithm can successfully parse a single packet sequence in to multiple packet sequences with no loss in website fingerprinting accuracy. Gu et al. [15] studied website fingerprinting in multi-tab browsing setting. Using a Naive Bayes classifier on the 50 top ranked websites in Alexa, they show when tabs are opened with a delay of 2 seconds, they can classify the first tab with 75.9% accuracy, and the background tab with 40.5%. More recently, Kwon et al. [19] showed that website fingerprinting attacks can be applied to Tor hidden services, and due to the long lived structure of hidden services, attacks can be even more accurate than when compared to non-hidden pages. They correctly deanonymize 50 monitored hidden service servers with TPR of 88% and FPR of 7.8% in an open world setting. We further improve on this in our work, resulting in a more accurate attack on the same data set.

---

[1] [17] considers an open world size of ∼35K but only tried to separate monitored pages from unmonitored pages instead of further classifying the monitored pages to the correct website. The authors assume the adversary monitors four pages: google.com, facebook.com, wikipedia.org and twitter.com. They trained a classifier using 36 traces for each of the Alexa Top 100 web pages, including the web pages of the monitored pages. The four traces for each of the monitored sites plus one trace for each of the unmonitored sites up to ∼35K are used for testing.

[2] All code will be made available through code repositories under a liberal open source license and data will be deposited in open data repositories.

In concurrent work, Panchenko et al. [28] have experimented with large datasets. Using a mix of different sources they produced two datasets, one of 34,580 unique websites (118,884 unique web pages) and another of 65,409 unique websites (211,148 unique web pages). Using a variation of a sequence of cumulative summations of packet sizes in a traffic trace they show their attack, CUMUL, was of similar accuracy to $k$-NN [39] under normal browsing conditions. However, we show that due to their feature set dependency on order and packet counting, their attack suffers substantially under simple website fingerprinting defenses and is outperformed by our technique, $k$-fingerprinting.

**Random Forests.** Random forests are a classification technique consisting of an ensemble of decision trees, taking a consensus vote of how to classify a new object. They have been shown to perform well in classification, regression [6, 20] and anomaly detection [22]. Each tree in the forest is trained using labeled objects represented as feature vectors of a fixed size. Training includes some randomness to prevent over-fitting: the training set for each tree is sampled from the available training set with replacement. Due to the bootstrap sampling process there is no need for $k$-fold cross validation to measure $k$-fingerprinting performance, it is estimated via the unused training samples on each tree [6]. This is referred to as the *out-of-bag* score.

## 3 Attack Design

We consider an attacker that can passively collect a client's encrypted or anonymized web traffic, and aims to infer which web resource is being requested. Dealing with an open-world, makes approaches based purely on classifying previously seen websites inapplicable. Our technique, $k$-fingerprinting, aims to define a distance-based classifier. It automatically manages unbalanced sized classes and assigns meaningful distances between packet sequences, where close-by 'fingerprints' denote requests likely to be for the same resources.

### 3.1 $k$-fingerprints from random forests

In this work we use random forests to extract a fingerprint for each traffic instance, instead of using directly the classification output of the forest. We define a distance metric between two traces based on the output of the forest: given a feature vector each tree in the forest associates a leaf identifier with it, forming a vector of leaf identifiers for the item, which we refer to as the *fingerprint*.

Once fingerprint vectors are extracted for two traces, we use the Hamming[3] distance to calculate the distance between these fingerprints[4]. We classify a test instance as the label of the closest $k$ training instances via the Hamming distance of fingerprints – assuming all labels agree. We evaluate the effect of varying $k$, the number of fingerprints used for comparison, in Sections 7, 8 and 9.

This leafs vector output represents a robust fingerprint: we expect similar traffic sequences are more likely to fall on the same leaves than dissimilar traffic sequences. Since the forest has been trained on a classification task, traces from the same websites are preferentially aggregated in the same leaf nodes, and those from different websites kept apart.

We can vary the number of training instances $k$ a fingerprint should match, to allow an attacker to trade the true positive rate (TPR) for false positive rate (FPR). This is not possible using the direct classification of the random forest. By using a $k$ closest fingerprint technique for classification, the attacker can choose how they wish to decide upon final classification[5]. For the closed-world setting we do not need the additional fingerprint layer for classification, we can simply use the classification output of the random forest since all classes are balanced and our attack does not have to differentiate between False Positives and False Negatives. For the closed-world setting we measure the mean accuracy of the random forest.

---

[3] We experimented with using the Hamming, Euclidean, Mahalanobis and Manhattan distance functions and found Hamming to provide the best results.

[4] For example, given the Hamming distance function $d : V \times V \to \mathbb{R}$, where $V$ is the space of leaf symbols, we expect given two packet sequences generated from loading *google.com*, with fingerprints vectors $f_1$, $f_2$ and a packet sequence generated from loading *facebook.com* with fingerprint $f_3$, that $d(f_1, f_2) < d(f_1, f_3)$ and $d(f_1, f_2) < d(f_2, f_3)$.

[5] We chose to classify a traffic instance as a monitored page if all $k$ fingerprints agree on the label, but an attacker could choose some other metric such as majority label out of the $k$ fingerprints.

## 3.2 The *k*-fingerprinting attack

Our *k*-fingerprinting attack proceeds in two phases: The attacker chooses which web pages they wish to monitor and captures network traffic generated via loading the monitored web pages and a large number of unmonitored web pages. These target traces for monitored websites, along with some traces for unmonitored websites, are used to train a random forest for classification. Given a packet sequence representing each training instance of a monitored web page, it is converted to a fixed length fingerprint as described in Section 3.1 and stored.

The attacker then passively collects instances of web page loads from a client's browsing session. A fingerprint is extracted from the newly collected packet sequence. The attacker then computes the Hamming distance of this new fingerprint against the corpus of fingerprints collected during training and is classified as a monitored page if and only if all *k* fingerprints agree on classification, as described in Section 3.1, otherwise it is classified as an unmonitored page.

We define the following performance measures for the attack:
- **True Positive Rate (TPR).** The probability that a monitored page is classified as the correct monitored page.
- **False Positive Rate (FPR).** The probability that an unmonitored page is incorrectly classified as a monitored page.
- **Bayesian Detection Rate (BDR).** The probability that a page corresponds to the correct monitored page given that the classifier recognized it as that monitored page. Assuming a uniform distribution of pages BDR can be found from TPR and FPR using the formula

$$\frac{TPR \cdot \Pr(M)}{(TPR \cdot \Pr(M) + FPR \cdot \Pr(U))}$$

where

$$\Pr(M) = \frac{|\text{Monitored}|}{|\text{Total Pages}|}, \quad \Pr(U) = 1 - P(M).$$

Ultimately BDR indicates the practical feasibility of the attack as it measures the main concern of the attacker, the probability that the classifier made a correct prediction.

## 4 Data gathering

We collect two data sets: one via Tor, $DS_{Tor}$, and another via a standard web browser, $DS_{Norm}$. $DS_{Norm}$ consists of 30 instances from each of 55 monitored web pages, along with 7,000 unmonitored web pages chosen from Alexas top 20,000 web sites [1]. We collected $DS_{Norm}$ using a number of `Amazon EC2 instances`[6], `Selenium`[7] and the headless browser `PhantomJS`[8]. We used `tcpdump`[9] to collect network traces for 20 seconds with 2 seconds between each web page load. Monitored pages were collected in batches of 30 and unmonitored web pages were collected successively. Page loading was performed with no caches and time gaps between multiple loads of the same web page, as recommended by Wang and Goldberg [38]. We chose to monitor web pages from Alexa's top 100 web sites [1] to provide a comparison with the real world censored web pages used in the Wang et al. [39] data set[10]. $DS_{Tor}$ was collected in a similar manner to $DS_{Norm}$ but was collected via the Tor browser. $DS_{Tor}$ consists of two subsets of monitored web pages: (i) 100 instances from each of the 55 top Alexa monitored web pages and (ii) 80 instances from each of 30 popular Tor hidden services[11]. The unmonitored set is comprised of the top 100,000 Alexa web pages, excluding the top 55. We chose to fingerprint web pages as listed by Alexa as these constitute the most popular web pages in the world over extended periods of time, and hence provide a more realistic dataset than choosing pages at random and/or using transiently popular website links as included in Panchenko et al.'s recent work [28]. By including website visits to trending topics we argue that this diminishes the ability to properly measure how effective a website fingerprinting attack will perform in general.

For comparison to previous work, we evaluated our attack on one of the previous largest website fingerprinting data sets [39], which collected 90 instances from each of 100 monitored sites, along with 5000 unmonitored web pages. The Wang et al. monitored web pages are various real-world censored websites from UK, Saudi Arabia and China providing a realistic set of web pages an attacker may wish to monitor. The unmonitored web pages are chosen at random from Alexa's top 10,000 websites – with no intersection between monitored and unmonitored web pages.

This allows us to validate *k*-fingerprinting on two different data sets while allowing for direct comparison against the state-of-the-art *k*-Nearest Neighbor attack [39]. We can also infer how well the attack works on censored web pages which may not have small landing pages or be set up for caching like websites in the top

---
[6] https://aws.amazon.com/ec2/
[7] http://www.seleniumhq.org/
[8] http://phantomjs.org/
[9] http://www.tcpdump.org/
[10] We used TCP/IP packets for final classification over abstracting to the Tor cell layer [38], preliminary experiments showed no consistent improvements from using one data layer for classification over the other.
[11] A Tor hidden service is a website that is hosted on a Tor client's *Onion Proxy*, which serves as the interface between application and network. Tor hidden services allow both a client accessing the website and the server hosting the website to remain anonymous to one another and any external observers. We chose hidden services to fingerprint based on popularity as listed by the .onion search engine http://www.ahmia.fi/.

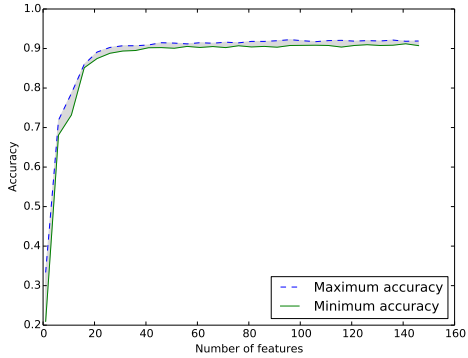

Figure 1: Accuracy of *k*-fingerprinting in a closed-world setting as the number of features is varied.

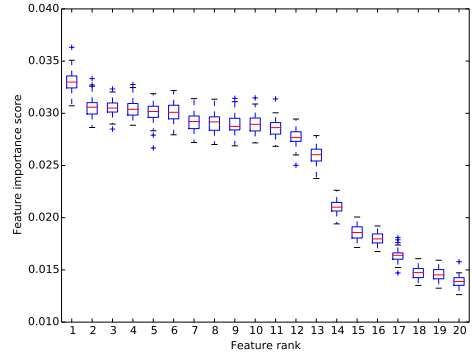

| № | Feature Description |
|---|---|
| 1. | Number of incoming packets. |
| 2. | Number of outgoing packets as a fraction of the total number of packets. |
| 3. | Number of incoming packets as a fraction of the total number of packets. |
| 4. | Standard deviation of the outgoing packet ordering list. |
| 5. | Number of outgoing packets. |
| 6. | Sum of all items in the alternative concentration feature list. |
| 7. | Average of the outgoing packet ordering list. |
| 8. | Sum of incoming, outgoing and total number of packets. |
| 9. | Sum of alternative number packets per second. |
| 10. | Total number of packets. |
| 11-18. | Packet concentration and ordering features list. |
| 19. | The total number of incoming packets stats in first 30 packets. |
| 20. | The total number of outgoing packets stats in first 30 packets. |

Figure 2: The 20 most important features.

Alexa list. Testing *k*-fingerprinting on both real-world censored websites and top alexa websites indicates how the attack performs across a wide range of websites.

For the sake of comparison, according to a study by research firm Nielsen [3] the number of unique websites visited per month by an average client in 2010 was 89. Another study [17, 26] collected web site statistics from 80 volunteers in a virtual office environment. Traffic was collected from each volunteer for a total of 40 hours. The mean unique number of websites visited per volunteer was 484, this is substantially smaller than the world sizes we consider in our experiments.

## 5 Feature selection

Our first contribution is a systematic analysis of feature importance. Despite some preliminary work by Panchenko et al. [27], there is a notable absence of feature analysis in the website fingerprinting literature. Instead features are picked based on heuristic arguments. All feature importance experiments were performed with the Wang et al. data set [39] so as to allow direct comparison with their attack results.

We train a random forest classifier in the closed-world setting using a feature vector comprised of features in the literature, and labels corresponding to the monitored sites. We use the gini coefficient as the purity criterion for splitting branches and estimate feature importance using the standard methodology described by Breiman [2, 6, 13]. Each time a decision tree branches on a feature the weighted sum of the gini impurity index for the two descendant nodes is higher than the purity of the parent node. We add up the gini decrease for each individual feature over the entire forest to get a consistent measure of feature importance.

Figure 1 illustrates the effect of using a subset of features for random forest classification. We first train a random forest classifier to establish feature importance; and then train new random forests with only subsets of the most informative features in batches of five. As we increase the number of features we observe a monotonic increase in accuracy; however there are diminishing returns as we can achieve nearly the same accuracy after using the 30 most important features. We chose to use 150 features in all following experiments since the difference in training time when using fewer features was negligible.

Figure 2 identifies the top-20 ranked features and illustrates their variability across 100 repeated experiments. As seen in Figure 1 there is a reduction in gradient when combining the top 15 features compared to using the top 10 features. Figure 2 shows that the top 13 features are comparatively much more important than the rest of the top 20 features, hence there is only a slight increase in accuracy when using the top 15 features compared to using the top 10. After the drop between the rank 13 and rank 14 features, feature importance falls steadily until feature rank 40, after which the reduction in feature importance is less prominent[12]. Note that there is some interchangeability in rank between features, we assign ranks based on the average rank of a feature over the 100 experiments.

---

[12] The total feature importance table is shown in Appendix 13.

**Feature Importance**

From each packet sequence we extract the following features:

- **Number of packets statistics.** The total number of packets, along with the number of incoming and outgoing packets [12, 27, 39]. The number of incoming packets during transmission is the most important feature, and together with the number of outgoing packets during transmission are always two of the five most important features. The total number of packets in transmission has rank 10.
- **Incoming & outgoing packets as fraction of total packets.** The number of incoming and outgoing packets as a fraction of the total number of packets [27]. Always two of the five most important features.
- **Packet ordering statistics.** For each successive incoming and outgoing packet, the total number of packets seen before it in the sequence [7, 27, 39]. The standard deviation of the outgoing packet ordering list has rank 4, the average of the outgoing packet ordering list has rank 7. The standard deviation of the incoming packet ordering list has rank 12 and the average of the incoming packet ordering list has rank 13.
- **Concentration of outgoing packets.** The packet sequence split into non-overlapping chunks of 20 packets. Count the number of outgoing packets in each of the chunks. Along with the entire chunk sequence, we extract the standard deviation (rank 16), mean (rank 11), median (rank 64) and max (rank 65) of the sequence of chunks. This provides a snapshot of where outgoing packets are concentrated [39]. The features that make up the concentration list are between the 15$^{th}$ and 30$^{th}$ most important features, but also make up the bulk of the 75 least important features.
- **Concentration of incoming & outgoing packets in first & last 30 packets.** The number of incoming and outgoing packets in the first and last 30 packets [39]. The number of incoming and outgoing packets in the first thirty packets has rank 19 and 20, respectively. The number of incoming and outgoing packets in the last thirty packets has rank 50 and 55, respectively.
- **Number of packets per second.** The number of packets per second, along with the mean (rank 44), standard deviation (rank 38), min (rank 117), max (42), median (rank 50).
- **Alternative concentration features.** This subset of features is based on the concentration of outgoing packets feature list. The outgoing packets feature list split into 20 evenly sized subsets and sum each subset. This creates a new list of features. Similarly to the concentration feature list, the alternative concentration feature list are regularly in the top 20 most important features and bottom 50 features. Note though concentration features are never seen in the top 15 most important features whereas alternative concentration features are, – at rank 14 and 15, – so information is gained by summing the concentration subsets.
- **Packet inter-arrival time statistics.** For the total, incoming and outgoing packet streams extract the lists of inter-arrival times between packets. For each list extract the max, mean, standard deviation, and third quartile [5]. These features have rank between 40 and 70.
- **Transmission time statistics.** For the total, incoming and outgoing packet sequences we extract the first, second, third quartile and total transmission time [39]. These features have rank between 30 and 50. The total transmission time for incoming and outgoing packet streams are the most important out of this subset of features.
- **Alternative number of packets per second features.** For the number of packets per second feature list we create 20 even sized subsets and sum each subset. The sum of all subsets is the 9$^{th}$ most important feature. The features produced by each subset are in the bottom 50 features - with rank 101 and below. The important features in this subset are the first few features with rank between 66 and 78, that are calculated from the first few seconds of a packet sequence.

We conclude that the total number of incoming packets is the most informative feature. This is expected as different web pages have different resource sizes, that are poorly hidden by encryption or anonymization. The number of incoming and outgoing packets as a fraction of the total number of packets are also informative for the same reason.

The least important features are from the padded concentration of outgoing packets list, since the original concentration of outgoing packets lists were of non-uniform size and so have been padded with zeros to give uniform length. Clearly, if most packet sequences have been padded with the same value this will provide a poor criterion for splitting, hence being a feature of low importance. Packet concentration statistics, while making up the bulk of "useless features" also regularly make up a few of the top 30 most important features, they are the first few items that are unlikely to be zero. In other words, the first few values in the packet concentration list do split the data well.

Packet ordering features have rank 4, 7, 12 and 13, indicating these features are a good criterion for classification. Packet ordering features exploit the information leaked via the way in which browsers request resources and the end server orders the resources to be sent. This supports conclusions in [7, 39] about the importance of packet ordering features.

We also found that the number of incoming and outgoing packets in the first thirty packets, with rank 19 and 20, were more important than the number of incoming and outgoing packets in the last thirty packets, with rank 50 and 55. In the alternative number packets per second feature list the earlier features were a better criterion for splitting than the later features in the list. This supports claims by Wang et al. [39] that the beginning of a packet sequence leaks more information than the end of a packet sequence. In contrast to Bissias et al. [5] we found packet inter-arrival time statistics, with rank between 40 and 70, only slightly increase the attack accuracy, despite being a key feature in their work.

## 6 Attack on hardened defenses

For direct comparison we tested our random forest classifier in a closed-world setting on various defenses against the $k$-NN attack and the more recent CUMUL [28] attack using the Wang et al. data set [39]. Note that most of these defenses require large bandwidth overheads that may render them unusable for the average client. We test against the following defenses:

- **BuFLO [12].** This defense sends packets at a constant size during fixed time intervals. This potentially extends the length of transmission and requires dummy packets to fill in gaps.
- **Decoy pages [27].** This defense loads a decoy page whenever another page is loaded. This provides background noise that degrades the accuracy of an attack. This is essentially a defense that employs multi-tab browsing.
- **Traffic morphing [40].** Traffic morphing shapes a clients traffic to look like another set of web pages. A client chooses the source web pages that they would like to defend, as well as a set of target web pages that they would like to make the source processes look like.
- **Tamaraw [35].** Tamaraw operates similarly to BuFLO but fixes packet sizes depending on their direction. Outgoing traffic is fixed at a higher packet interval, this reduces overhead as outgoing traffic is less frequent.
- **Adaptive Padding (AP) [18, 31].** AP protects anonymity by introducing traffic in to statistically unlikely delays between packets in a flow. This limits the amount of extra bandwidth required and does not incur any latency costs. AP uses previously computed histograms of inter-arrival packet times from website loads to determine when a dummy packet should be injected[13]. This is currently the favored option if padding were to be implemented in Tor [4].

[13] As Juarez et al. [18] note, the distribution of histogram bins is dependent on the individual client bandwidth capacity. Optimizing histograms for a large number of clients is an open problem. Here we implement a naive version of AP with one master histogram for all clients.

Table 1: Attack comparison under various website fingerprinting defenses.

| Defenses | This work | $k$-NN [39] | CUMUL [28] | Bandwidth overhead (%) |
|---|---|---|---|---|
| No defense | 0.91 ±0.01 | 0.91 ±0.03 | 0.91 ±0.04 | 0 |
| Morphing [40] | **0.90** ±0.03 | 0.82 ±0.06 | 0.75 ±0.07 | 50 ±10 |
| Decoy pages [27] | **0.37** ±0.01 | 0.30 ±0.06 | 0.21 ±0.02 | 130 ±20 |
| Adaptive Padding [31] | **0.30** ±0.04 | 0.19 ±0.03 | 0.16 ±0.03 | 54 |
| BuFLO [12] | **0.21** ±0.02 | 0.10 ±0.03 | 0.08 ±0.03 | 190 ±20 |
| Tamaraw [35] | **0.10** ±0.01 | 0.09 ±0.02 | 0.08 ±0.03 | 96 ±9 |

Table 1 shows the performance of $k$-fingerprinting against $k$-NN and CUMUL under various website fingerprinting defenses in a closed-world setting. Under every defense $k$-fingerprinting is comparable or achieves better results than the $k$-NN attack and performs significantly better than CUMUL. Note that $k$-fingerprinting does equally well when traffic morphing is applied compared to no defense. As Lu et al. [23] note, traffic morphing is only effective when the attacker restricts attention to the same features targeted by the morphing process. Our results confirm that attacks can succeed even when traffic morphing is employed. $k$-fingerprinting also performs nearly 10% better than $k$-NN when decoy pages are used, which is in effect a marker for how well the attack performs on multi-tab browsing. Due to the dependency of packet length and sequence length features, CUMUL performs substantially worse than the other two attacks under website fingerprinting defenses. Though CUMUL uses a similar number of features and is of similar computational efficiency to $k$-fingerprinting, simple defenses remove the feature vector patterns between similar web pages used in CUMUL, rendering the attack ineffectual. More generally, any attack that uses a restricted set of features will suffer greatly from a defense that targets those features. $k$-fingerprinting performs well under defenses due to its feature set that captures traffic information not used in CUMUL such as packet timings and burst patterns. The $k$-NN attack performs slightly better than CUMUL but requires an order of magnitude more features than both CUMUL and $k$-fingerprinting. Our attack is both more efficient and more accurate than CUMUL and $k$-NN under defenses.

## 7 $k$-fingerprinting the Wang et al. data set

We first evaluate $k$-fingerprinting on the Wang et al. data set [39]. This data set was collected over Tor, and thus implements padding of packets to fixed-size cells (512-bytes) and randomization of request orders [30]. Thus the only available information to $k$-fingerprinting are timing and volume features. We train on 60 out of the 90 instances for each of the 100 monitored web pages; we vary the number of pages on which we train from the 5000 unmonitored web pages. For the attack evaluation we use fingerprints of length 200 and 150 features. Final classification is as described in Section 3.2, if all $k$ finger-prints agree on classification a test instance is classified as a monitored web page, otherwise it is classified as an unmonitored web page.

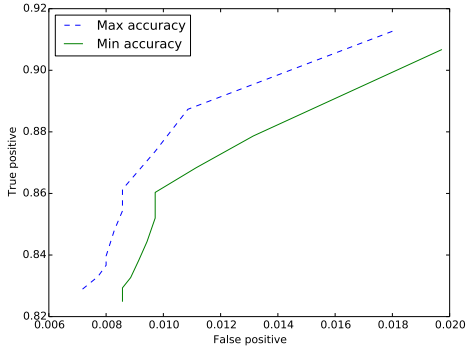
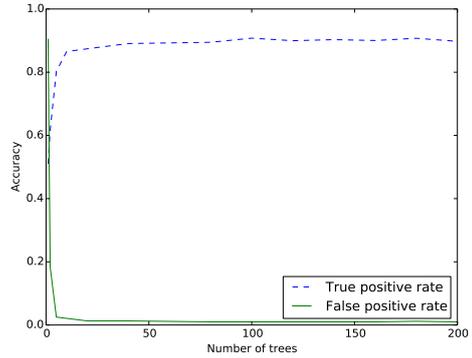

Figure 3: Attack results for 1500 unmonitored training pages while varying the number of fingerprints used for comparison, *k*, over 10 experiments.

Figure 4: Accuracy of *k*-fingerprinting as we vary the number of trees in the forest.

Table 2: *k*-fingerprinting results for *k*=3 while varying the number of unmonitored training pages.

| Training pages | TPR | FPR | BDR |
|---:|---|---|---|
| 0 | $0.90 \pm 0.02$ | $0.750 \pm 0.010$ | 0.419 |
| 1500 | $0.88 \pm 0.02$ | $0.013 \pm 0.007$ | 0.983 |
| 2500 | $0.88 \pm 0.01$ | $0.007 \pm 0.001$ | 0.993 |
| 3500 | $0.88 \pm 0.01$ | $0.005 \pm 0.001$ | 0.997 |
| 4500 | $0.87 \pm 0.02$ | $0.009 \pm 0.001$ | 0.998 |

The scenario for the attack is as follows: an attacker monitors 100 web pages; they wish to know whether a client is visiting one of those pages, and establish which one. The client can browse to any of these web pages or to 5000 unmonitored web pages, which the attacker classifies in bulk as an unmonitored page.

Using the *k*-fingerprinting method for classifying a web page we measure a TPR of $0.88 \pm 0.01$ and a FPR of $0.005 \pm 0.001$ when training on 3500 unmonitored web pages and *k*, the number of training instances used for classification, set at *k*=3. *k*-fingerprinting achieves better accuracy than the state-of-the-art *k*-NN attack that has a TPR of $0.85 \pm 0.04$ and a FPR of $0.006 \pm 0.004$. Given a monitored web page *k*-fingerprinting will misclassify this page 12% of the time, while *k*-NN will misclassify with 15% probability.

Best results are achieved when training on 3500 unmonitored web pages. Table 2 reports TPR and FPR when using different numbers of unmonitored web pages for training with *k*=3. As we train more unmonitored web pages we decrease our FPR with almost no reduction in TPR. After training 3500 unmonitored pages there is no decrease in FPR and so no benefit in training more unmonitored web pages. This is confirmed by the marginal increase in BDR after training on at least some of the unmonitored set. Furthermore without training on any of the unmonitored web pages, despite the high FPR the classifier has more than 40% probability of being correct when classifying a web page as monitored.

Figure 3 illustrates how classification accuracy changes as, *k*, the number of fingerprints used for classification changes. For a low *k* the attack achieves a FPR of around 1%, as we increase the value of *k* we reduce the number of misclassifications since it is less likely that all *k* fingerprints will belong to the same label, but we also reduce the TPR. Altering the number of fingerprints used for classification allows an attacker to tune the classifier to either a low FPR or high TPR depending on the desired application of the attack.

We find that altering the number of fingerprints used for classification, *k*, affects the TPR and FPR more than the number of unmonitored training pages. This suggests that while it is advantageous to have a large world size of unmonitored pages, increasing the number of unmonitored training pages does not increase accuracy of the classifier dramatically. This supports Wang et al.'s [39] claims to the same effect. In practice an attacker will need to train on at least some unmonitored pages to increase the BDR, though this does not need to be a substantial amount; training 1500 unmonitored web pages leads to a 98.3% chance the classifier is correct when claiming to have recognized a monitored web page.

**Fingerprint length.** Changing the length of the fingerprint vector will affect *k*-fingerprinting accuracy. For a small fingerprint length there may not be enough diversity to provide an accurate measure of distance over all packet sequences. Figure 4 shows the resulting TPR and FPR as we change the length of fingerprints in the Wang et al. [39] data set. We set *k*=1 and train on 4000 unmonitored web pages. Using only a fingerprint of length one results in a TPR of 0.51 and FPR of 0.904. Clearly us-ing a fingerprint of length one results in a high FPR since there is a small universe of leaf symbols from which to create the fingerprint. A fingerprint of length 20 results in a TPR of 0.87 and FPR of 0.013. After this there are diminishing returns for increasing the length of the fingerprint vector.

Table 3: Attack results on top Alexa sites for *k*=2 while varying the number of unmonitored training pages.

| Training pages | TPR | FPR | BDR |
|---:|---:|---:|---:|
| 2000 | $0.93 \pm 0.03$ | $0.032 \pm 0.010$ | 0.33 |
| 4000 | $0.93 \pm 0.01$ | $0.018 \pm 0.007$ | 0.47 |
| 8000 | $0.92 \pm 0.01$ | $0.008 \pm 0.002$ | 0.67 |
| 16000 | $0.91 \pm 0.02$ | $0.003 \pm 0.001$ | 0.86 |

Table 4: Attack results on Tor hidden services for *k*=2 while varying the number of unmonitored training pages.

| Training pages | TPR | FPR | BDR |
|---:|---:|---:|---:|
| 2000 | $0.82 \pm 0.03$ | $0.0020 \pm 0.0015$ | 0.72 |
| 4000 | $0.82 \pm 0.04$ | $0.0007 \pm 0.0006$ | 0.88 |
| 8000 | $0.82 \pm 0.02$ | $0.0002 \pm 0.0001$ | 0.96 |
| 16000 | $0.81 \pm 0.02$ | $0.0002 \pm 0.0002$ | 0.97 |

## 8 Attack evaluation on $DS_{Tor}$

We now evaluate *k*-fingerprinting on $DS_{Tor}$. First we evaluate the attack given a monitored set of the top 55 Alexa web pages, with 100 instances for each web page. Then we evaluate the attack given a monitored set of 30 Tor hidden services, with 80 instances for each hidden service. The unmonitored set remains the same for both evaluations, the top 100,000 Alexa web pages with one instance for each web page.

### 8.1 Alexa web pages monitored set

Table 3 shows the accuracy of *k*-fingerprinting as the number of unmonitored training pages is varied. For the monitored web pages, 70 instances per web page were trained upon and testing was done on the remaining 30 instances of each web page. As expected, the FPR decreases as the number of unmonitored training samples grows. Similar to Section 7 there is only a marginal decrease in TPR while we see a large reduction in the FPR as the number of training samples grows. Meaning an attacker will not have to compromise on TPR to decrease the FPR; when scaling the number of unmonitored training samples from 2% to 16% of the entire set the TPR decreases from 93% to 91% while the FPR decreases from 3.2% to 0.3%. There is a more pronounced shift in BDR with the increase of unmonitored training pages, however an attacker needs to train on less than 10% of the entire dataset to have nearly 70% confidence that classifier was correct when it claims to have detected a monitored page.

Clearly the attack will improve as the number of training samples grows, but in reality an attacker may have limited resources and training on a significant fraction of 100,000 web pages may be unfeasible. Figure 5 shows the TPR and FPR of *k*-fingerprinting as the number of unmonitored web pages used for testing grows while the number of unmonitored web pages used for training is kept at 2000, for different values of *k*. We may think of this as the evaluation of success of *k*-fingerprinting as a client browses to more and more web pages over multiple browsing sessions. Clearly for a small *k*, both TPR and FPR will be comparatively high. Given that, with *k*=5 only 2.5% of unmonitored web pages are falsely identified as monitored web pages, out of 98,000 unmonitored web pages.

### 8.2 Hidden services monitored set

Table 4 shows the accuracy of *k*-fingerprinting as the number of unmonitored training pages is varied. For the monitored set, 60 instances per hidden service were trained upon and testing was done on the remaining 20 instances of each hidden service. Again we observe a marginal loss of TPR and a large reduction in FPR as the number of training samples grows. When scaling the number of unmonitored training samples from 2% to 16% of the entire set the TPR decreases from 82% to 81% while the FPR decreases by an order of magnitude from 0.2% to 0.02%. As a result, when training on 16% of the unmonitored set only 16 unmonitored web pages out of 84,000 were misclassified as a Tor hidden service. In comparison to the Alexa web pages monitored set the TPR is around 10% lower, while the FPR is also greatly reduced. This is evidence that Tor hidden services are easy to distinguish from standard web pages loaded over Tor. There is also a higher but more gradual increase in BDR compared to standard web pages. An attacker need only train on as little as 2% of unmonitored pages to have over 70% confidence that classification of a monitored page was correct, with this rising to 97% when training on 16% of the unmonitored dataset.

Similarly to Figure 5, Figure 6 shows the TPR and FPR of *k*-fingerprinting as the number of unmonitored web pages used for testing grows while the number of unmonitored web pages used for training is kept at 2000, for different values of *k*. Both the TPR and FPR is lower than in Figure 5. For example using *k*=5, the FPR is 0.2% which equates to only 196 out of 98,000 unmonitored pages being falsely classified as monitored pages.

From Figure 7 we observe that the BDR of both standard web pages and hidden services monitored sets depends heavily on not only the world size but the number of fingerprints used for classification. With *k*=10, when a web page is classified as a monitored hidden service page, there is over an 80% chance that the classifier was correct, despite the unmonitored world size (98,000) being over 160 times larger than the monitored world size (600). The high BDR regardless of the disparity in world sizes makes it clear that our attack is highly effective under realistic large world size conditions.

It is clear that an attacker need only train on a small fraction of data to launch a powerful fingerprinting attack. It is also clear that Tor hidden services are easily distinguished from standard web pages, rendering them vulnerable to website fingerprinting attacks. We attribute the lower FPR of Tor hidden services when compared to a monitored training set of standard web page traffic to this distinguishability. A standard web page over Tor is more likely to be confused with another standard web page than a Tor hidden service.

**Comparison with Kwon et al. [19] hidden services results**. For comparison we ran $k$-fingerprinting on the data set used in the Kwon et al. study on fingerprinting hidden services. This data set simulated a client connecting to a hidden service. The data set consists of 50 instances for each of 50 monitored hidden services and an unmonitored set of 950 hidden services. When training on 100 of the unmonitored pages they report attack accuracy of 0.9 TPR and 0.4 FPR. $k$-fingerprinting achieved a similar true positive rate but the FPR is reduced to 0.22. This FPR reduction in comparison with Kwon et al. continued regardless of the amount of data used for training.

## 9 Attack evaluation on $DS_{Norm}$

Besides testing on $DS_{Tor}$, Wang et al. [39] data set and the Kwon et al. [19] data set we tested the efficacy of $k$-fingerprinting on $DS_{Norm}$. This allows us to establish how accurate $k$-fingerprinting is over a standard encrypted web browsing session or through a VPN.

### 9.1 Attack on encrypted browsing sessions

An encrypted browsing session does not pad packets to a fixed size so the attacker may extract the following features in addition to time features:
- **Size transmitted.** For each packet sequence we extract the total size of packets transmitted, in addition, we extract the total size of incoming packets and the total size of outgoing packets.
- **Size transmitted statistics.** For each packet sequence we extract the average, variance, standard deviation and maximum packet size of the total sequence, the incoming sequence and the outgoing sequence.

Apart from this modification in available features, the attack setting is similar: An attacker monitors a client browsing online and attempts to infer which web pages they are visiting. The only difference is that browsing with the Transport Layer Security (TLS) protocol, or Secure Sockets Layer (SSL) protocol, versions 2.0 and 3.0, exposes the destination IP address and port. The attack is now trying to infer which web page the client is visiting from the known website[14].

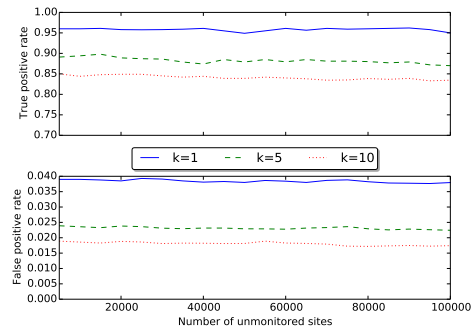

Figure 5: Attack accuracy on $DS_{Tor}$ with Alexa monitored set.

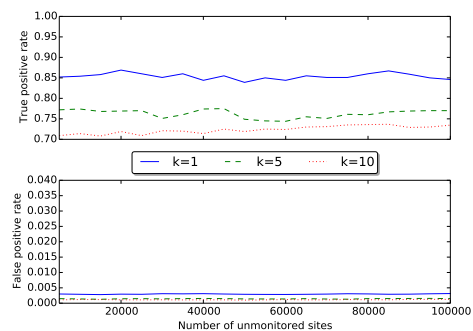

Figure 6: Attack accuracy on $DS_{Tor}$ with Tor hidden services monitored set.

The attacker monitors 55 web pages; they wish to know if the client has visited one of these pages. The client can browse to any of these web pages or to 7000 other web pages, which the attacker does not care to classify other than as unmonitored. We train on 20 out of the 30 instances for each monitored page and vary the number of unmonitored pages on which we train.

Despite more packet sequence information to exploit, the larger cardinality of world size gives rise to more opportunities for incorrect classifications. The attack achieves a TPR of 0.87 and a FPR of 0.004. We achieved best results when training on 4000 unmonitored web pages. Table 5 reports results for training on different numbers of unmonitored web pages, with $k = 2$. Figure 8 shows our results when modifying the number of fingerprints used ($k$) and training on 2000 unmonitored pages. We find that altering the number of unmonitored training pages decreases the FPR while only slightly decreasing the TPR. This mirrors our experimental findings from $DS_{Tor}$ and the Wang et al. data set.

---

[14]Note that the data sets are composed of traffic instances from some websites without SSL and TLS, as well as websites using the protocols. We expect our experiment conditions are much larger than the number possible web pages an attacker may wish to fingerprinting from a standard website.

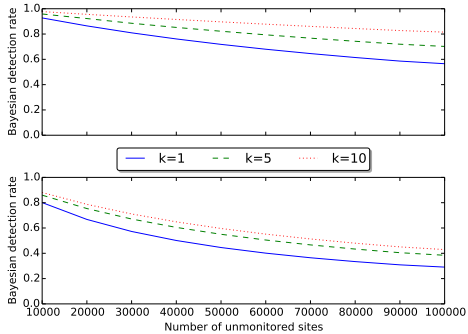

Figure 7: BDR for hidden services monitored set (above) and Alexa monitored set (below).

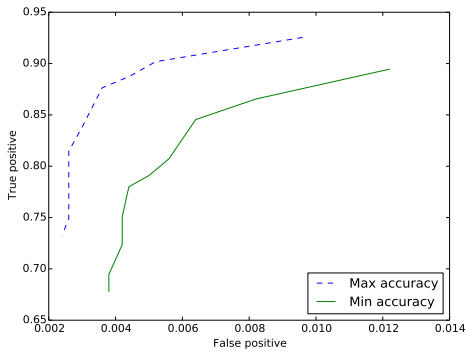

Figure 8: Attack results for 2000 unmonitored training pages while varying the number of fingerprints used for comparison, *k*, over 10 experiments.

### 9.2 Attack without packet size features

$DS_{Norm}$ was not collected via Tor and so also contains packet size information. We remove this to allow for comparison with $DS_{Tor}$ and the Wang et al. data set which was collected over Tor. This also gives us a baseline for how much more powerful *k*-fingerprinting is when we have additional packet size features available.
We achieved a TPR of 0.81 and FPR of 0.005 when training on 5000 unmonitored web pages. Table 6 shows our results at other sizes of training samples, with *k*=2. Removing packet size features reduces the TPR by over 0.05 and increases the FPR by 0.001. Clearly packet size features improve our classifier in terms of correct identifications but do not decrease the number of unmonitored test instances that were incorrectly classified as a monitored page. Despite the difference in FPR when including packet size information, the BDR is similar, suggest-ing that BDR is dominated by the amount of information that can be trained upon.

Table 5: Attack results with packet size features for a varying number of unmonitored training pages.

| Training pages | TPR | FPR | BDR |
|---:|---|---|---|
| 0 | $0.95 \pm 0.01$ | $0.850 \pm 0.010$ | 0.081 |
| 2000 | $0.90 \pm 0.01$ | $0.010 \pm 0.004$ | 0.908 |
| 4000 | $0.87 \pm 0.02$ | $0.004 \pm 0.001$ | 0.976 |
| 6000 | $0.86 \pm 0.01$ | $0.005 \pm 0.002$ | 0.990 |

Table 6: Attack results without packet size features for a varying number of unmonitored training pages.

| Training pages | TPR | FPR | BDR |
|---:|---|---|---|
| 0 | $0.90 \pm 0.01$ | $0.790 \pm 0.020$ | 0.082 |
| 2000 | $0.83 \pm 0.01$ | $0.009 \pm 0.001$ | 0.910 |
| 4000 | $0.81 \pm 0.02$ | $0.006 \pm 0.001$ | 0.961 |
| 6000 | $0.80 \pm 0.02$ | $0.005 \pm 0.001$ | 0.989 |

**Closed-World.** In the closed-world setting in which the client can only browse within the 55 monitored web pages *k*-fingerprinting is 0.91, compared to 0.96 when packet size features are available. In the closed-world setting attack accuracy improves by 5% when we include packet size features.

## 10 Fine grained open-world false positives on Alexa monitored set of $DS_{Tor}$

We observe that the classification error is not uniform across all web pages[15]. Some pages are misclassified many times, and confused with many others, while others are never misclassified. An attacker can leverage this information to estimate the misclassification rate of each of the web page classes instead of using the global average misclassification rate. A naive approach to this problem would be to first find which fingerprints contribute to the many misclassifications and remove them. Our analysis shows that the naive approach of removing "bad" fingerprints that contribute to many misclassifications will ultimately lead to a higher misclassification rate. Figure 9 shows the 50 fingerprints that cause the most misclassifications, and also shows for those same fingerprints the number of correct classifications they contribute towards. As we can see nearly all "bad" fingerprints actually contribute to many correct classifications. One may think it may still be beneficial to remove these fingerprints as the cumulative sum of misclassifications outweigh the number of correct classifications. This removal will then promote fingerprints that are further away in terms of Hamming distance from the fingerprinting that is being tested,

---
[15]See additional evidence in Appendix 14.

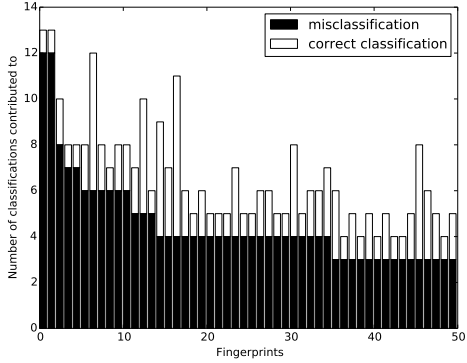

Figure 9: The fingerprints that lead to the most misclassifications and the correct classifications they contribute towards. Training on 2% of unmonitored pages with $k=3$.

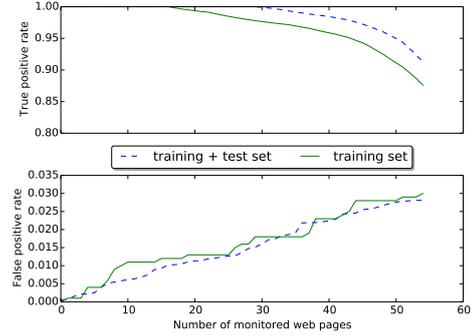

Figure 10: Rates for training on 1000 unmonitored pages, testing on 1000, and comparison when training on the full 2000 unmonitored pages and testing on the remaining 98000 unmonitored pages in $DS_{Tor}$, $k=3$.

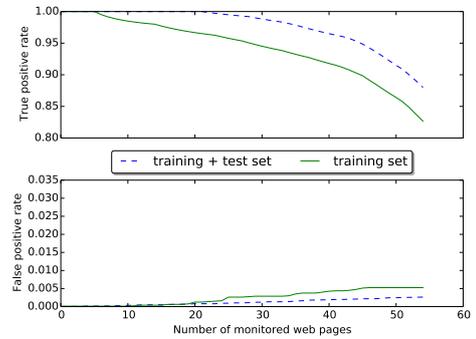

Figure 11: Rates for training on 8000 unmonitored pages, testing on 8000, and comparison when training on the full 16000 unmonitored pages and testing on the remaining 84000 unmonitored pages in $DS_{Tor}$, $k=3$.

which will lead to a greater number of misclassifications.

Instead an attacker can use their training set of web pages to estimate the TPR and FPR of each web page class, by splitting the training set in to a smaller training set and validation set. Since both sets are from the original training set the attacker has access to the true labels. The attacker then computes the TPR and FPR rates of each monitored class, this is used as an estimation for TPR and FPR when training on the entire training set and testing on new traffic instances. More specifically we split, for the monitored training set of 70 instance for each of the Alexa top 55 web pages, into smaller training sets of 40 instances and validation sets of 30 instances. This is used as a misclassification estimator for the full monitored training set against the monitored test set of 30 instances per class. Similarly we split the unmonitored training in half, one set as a smaller training set and the other as a validation set.

Figure 10 shows the TPR and FPR estimation accuracy for 2000 unmonitored training pages. Monitored classes are first ordered from best to worst in terms of their classification accuracy. Even with a small unmonitored training set of 2000 web pages, which is then split in to a training set of 1000 web pages and a validation set of 1000 web pages, an attacker can accurately estimate the FPR of the attack if some of the monitored classes were removed. For example, using only the best 20 monitored classes (in terms of TPR), an attacker would estimate that using those 20 classes as a monitored set, the FPR would be 0.012. Using the entire data set we see that the true FPR of these 20 classes is 0.010; the attacker has nearly precisely estimated the utility of removing a large fraction of the original monitored set.

There is a small difference between estimated and the actual FPR in both Figures 10 and 11. There is little benefit in training more unmonitored data if the attacker wants to accurately estimate the FPR; Figure 10 has a similar gap between the estimated FPR and true FPR when compared to Figure 11. It is evident even with a small training set, an attacker can identify web pages that are likely to be misclassified and then accurately calculate the utility of removing these web pages from their monitored set. Due to the overwhelmingly large world size of unmonitored web pages the BDR of Figure 10 does not grow dramatically with the removal of web pages that are likely to be misclassified; using the entire monitored set the BDR is 0.33, removing half of the monitored web pages the BDR is 0.35.

## 11 Attack Summary & Discussion

**Attack Summary.** Best attack results on data sets were achieved when training on approximately two thirds of the unmonitored web pages. Despite this, results from $DS_{Tor}$ show that an attacker can achieve a very small false positive rate while only training on 2% of the unmonitored data. Training on 2% of 100,000 unmoni-tored web pages greatly reduces the attack set up costs while only marginally reducing the accuracy compared to training on more data, providing a realistic setting under which an attack could be launched. Results on all data sets also suggest that altering *k*, the number of fingerprints used for classification, has a greater influence on accuracy than the number of training samples[16].

*k*-fingerprinting is robust; our technique achieves the same accuracy regardless of the type of monitored set or the manner in which it was collected (through Tor or standard web browsers). The monitored set in the Wang et al. [39] data set consists of real world censored websites, the Kwon et al. [19] monitored set consist of Tor hidden services and the $DS_{Tor/Norm}$ monitored sets were taken from Tor hidden services and top Alexa websites. We do see a reduction in FPR when the target monitored set are Tor hidden services due to the distinguishability between the hidden services and unmonitored standard web pages.

We also highlight the non-uniformity of classification performance: when a monitored web page is misclassified, it is usually misclassified on multiple tests. We show that an attacker can use their training set to estimate the error rate of *k*-fingerprinting per web page, and select targets with low misclassification rates.

**Computational Efficiency.** *k*-fingerprinting is more accurate and uses fewer features than state-of-the-art attacks. Furthermore *k*-fingerprinting is faster than current state-of-the-art website fingerprinting attacks. On the Wang et al. data set training time for 6,000 monitored and 2,500 unmonitored training pages is 30.738 CPU seconds on an 1.4 GHz Intel Core i5z. The *k*-NN attack [39] has training time per round of 0.064 CPU seconds for 2500 unmonitored training pages. For 6,000 rounds training time is 384.0 CPU seconds on an AMD Opteron 2.2 GHz cores. This can be compared to around 500 CPU hours using the attack described by Cai et al. [7]. Testing time per instance for *k*-fingerprinting is around 0.1 CPU seconds, compared to 0.1 CPU seconds to classify one instance for *k*-NN and 450 CPU seconds for the attack described by Cai et al. [7].

**Discussion.** Website fingerprinting research has been criticized for not being applicable to real-world scenarios [17, 29]. We have shown that a website fingerprinting attack can scale to the number of traffic instance an attacker may sample over long period of time with a high BDR and low FPR. However, we did not consider the cases where background traffic may be present, for example from multitab browsing, or the effect that short-lived websites will have on our attack. Gu et al. [15] show in their work that a simple Naive-Bayes attack achieves highly accurate results even when a client browses in multiple tabs. Wang and Goldberg [36] also show that website fingerprinting is effective in practical scenarios. With no prior attack set-up to tailor to a multi-tab browsing session our attack was able to classify nearly 40% of monitored pages correctly when the decoy defense was employed.

Website content rapidly changes which will negatively affect the accuracy of a website fingerprinting attack [17]. As the content of a website changes so will the generated packet sequences, if an attacker cannot train on this new data then an attack will suffer. However we note that an attack will suffer from the ephemeral nature of websites at different rates depending on the type of website being monitored. For example, an attack monitoring a news or social media site can expect a faster degradation in performance compared to an attack monitoring a landing page of a top 10 Alexa site [1]. Also note Tor does not cache by default, so if in the realistic scenario where an attacker wanted to monitor *www.socialmediawebsite.com* a client would be forced to navigate to the social media website landing page, which is likely to host content that is long lived and not subject to change. The problem of content change is weakened when fingerprinting Tor hidden services. As show by Kwon et al. [19] hidden pages show minimal changes in comparison to non-hidden pages, resulting in devastatingly accurate attacks on hidden services that can persist.

## 12 Conclusion

We establish that website fingerprinting attacks are a serious threat to online privacy. Clients of both Tor and standard web browsers are at risk from website fingerprinting attacks regardless of whether they browse to hidden services or standard websites. *k*-fingerprinting improves on state-of-the-art attacks in terms of both speed and accuracy: current website fingerprinting defenses either do not defend against *k*-fingerprinting or incur very high bandwidth overheads. Our world size is an order of magnitude larger than previous website fingerprinting studies, and twice as large in terms of unique website than Panchenko et al.'s recent work [28]. We have validated our attack on four separate datasets showing that it is robust and not prone to overfit one dataset, and so is applicable to real world browsing environments at scale. *k*-fingerprinting is highly accurate even when an attacker trains on a small fraction of the total data. Untrustworthy data within that small fraction can then be filtered and removed before the attack is launched to later yield better results, showing that long term website fingerprinting attacks on a targeted client is a realistic threat.

## References

[1] Alexa The Web Information Company, [Accessed

---

[16]Figure 17 illustrates that compared to training on a small number of monitored instances increasing the size of the monitored training set only incrementally increases accuracy.

## 13 Total feature importance.

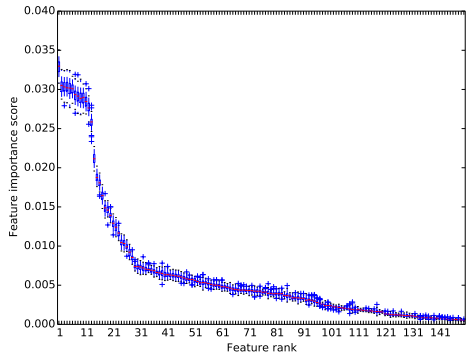

| № | Feature Description |
|---|---|
| 131 - 150 . | Packet concentration list features. |

Figure 12: Feature importance score for all 150 features in order. The table gives the description for the 20 least important features.

## 14 Closed World Error Rates

Figure 13 shows the confusion matrix in our closed-world setting, that is, it shows the 49 misclassifications (out of 550). We see that some persistent misclassification patterns of web pages appear, for example web page 54 is classified correctly four times but is misclassified as web page 0 six times. The misclassification rate in Figure 13 is 0.09 but this is the average error rate across all web pages.

Figure 13 shows that the classification error is not uniform across all web pages. Some pages are misclassified many times, and confused with many others, while others are never misclassified. An attacker can leverage this information to estimate the misclassification rate of each web page instead of using the global average misclassification rate.

As in Section 10, an attacker can use their training set of web pages to estimate the misclassification rate of each web page, by splitting the training set in to a smaller training set and validation set. Since both sets are from the original training set the attacker has access to the true labels. The attacker then computes the misclassification rate of each web page, which they can use as an estimation for the misclassification rate when training on the entire training set and testing on new traffic instances.

Figures 14 and 15 show the global misclassification rate for a varying number of monitored pages. Monitored pages are first ordered in terms of the misclassification rate they have, ordered from smallest to largest. From Figure 14, using the Wang et al. data set, we see that if the attacker considers only the top 50% on web pages in terms of per page misclassification rate, the true global misclassification rate and global misclassification rate estimated by the attacker drop by over 70%. Similarly from Figure 15, using $DS_{Norm}$, if the attacker considers only the top 50% on web pages in terms of per page misclassification rate, the true global misclassification rate and global misclassification rate estimated by the attacker drop by over 80%. This allows an attacker to train on monitored pages and then cull the pages that have too high an error rate, allowing for more confidence in the classification of the rest of the monitored pages.

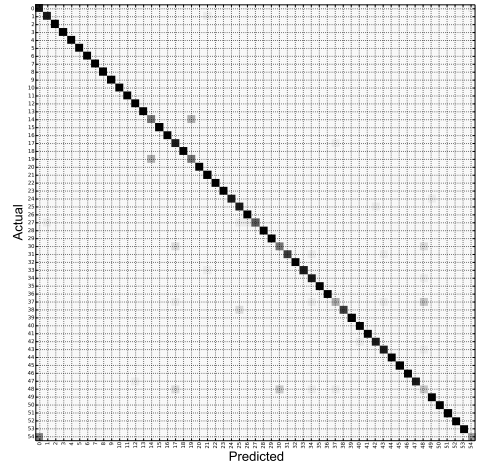

Figure 13: Confusion matrix for closed-world attack on Tor using $DS_{Norm}$. F1 score = 0.913, Accuracy: 0.915, 550 items.

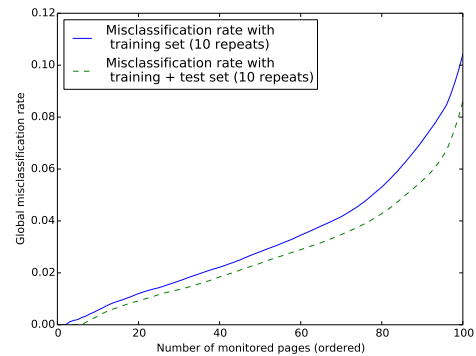

Figure 14: The global misclassification rate when considering different numbers of monitored pages from the Wang et al. data set. The monitored pages are ordered in terms of smallest misclassification rate to largest.

The gap between the attacker's estimate and the misclassification rate of the test set is largely due to the size of the data set. Figure 14 has a smaller error of estimate than Figure 15 because the Wang et al. data set has 60 instances per monitored page, in comparison $DS_{Norm}$ has 20 instances per monitored page. In practice, an attacker cannot expect perfect alignment; they are generated from two different sets of data, the training and training + test set. Nevertheless the attacker can expect this difference to decrease with the collection of more training instances.

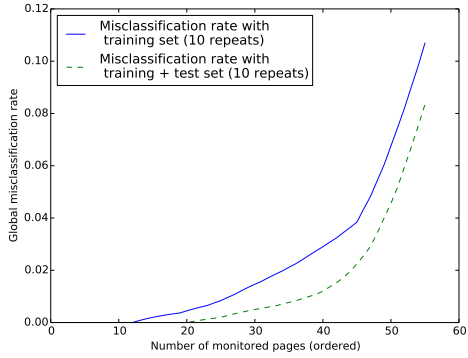

Figure 15: The global misclassification rate when considering different numbers of monitored pages from $DS_{Norm}$. The monitored pages are ordered in terms of smallest misclassification rate to largest.

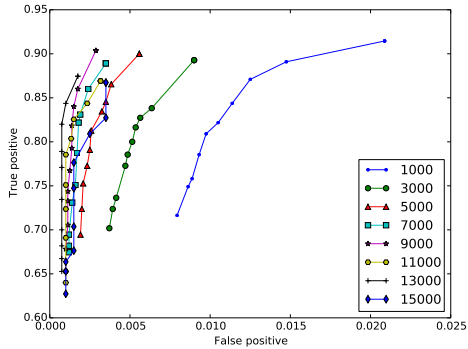

Figure 16: Attack accuracy for 17,000 unmonitored web pages. Each line represents a different number of unmonitored web pages that were trained, while varying k, the number of fingerprints used for classification.

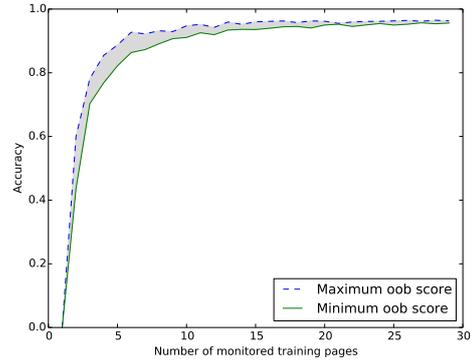

Figure 17: Attack out-of-bag score while varying the number of monitored training pages.

## 15 Attack on larger world size of $DS_{Norm}$

We run *k*-fingerprinting on $DS_{Norm}$ with the same number of monitored sites but increase the numbered of unmonitored sites to 17,000. We evaluate when we have both time and size features available.

Figure 16 shows the results of *k*-fingerprinting while varying the number of fingerprints (*k*) used for classification, from between 1 and 10, for various experiments trained with different numbers of unmonitored pages. We see that the attack results are comparable to the attack on 7000 unmonitored pages, meaning there is no degradation in attack accuracy when we increase the world size by 10,000 web pages. Training on approximately 30% of the 7000 unmonitored web pages *k*-fingerprinting gives a TPR of over 0.90 and a FPR of 0.01 for *k*=1. Training on approximately 30% of the 17,000 unmonitored web pages *k*-fingerprinting gives a TPR of 0.90 and a FPR of 0.006 for *k*=1.

The fraction of unmonitored pages that were incorrectly classified as a monitored page decreased as we increased our world size. In other words, out of 12,000 unmonitored pages only 72 were classified as a monitored page, with this Figure dropping to 24 if we use *k*=10 for classification. This provides a strong indication that *k*-fingerprinting can scale to a real-world attack in which a client is free to browse the entire internet, with no decrease in attack accuracy.

**Number of monitored training pages in closed-world.** Figure 17 shows the *out-of-bag* score as we change the number of *monitored* pages we train. We found that training on any more than a third of the data gives roughly the same accuracy.